\documentclass[%
 reprint,
superscriptaddress,
 amsmath,amssymb,
 aps,
pra,
]{revtex4-1}

\usepackage{graphicx}% Include figure files
\usepackage{epstopdf}
\usepackage{dcolumn}% Align table columns on decimal point
\usepackage{bm}% bold math
\usepackage{braket}
\usepackage{color}
\usepackage{ragged2e}
\usepackage{amsmath, bm}
\begin{document}
\preprint{APS/123-QED}

\title{Increasing two-photon entangled dimensions by shaping input beam profiles}

\author{Shilong Liu}
 \affiliation{Key Laboratory of Quantum Information, University of Science and Technology of China, Hefei, Anhui 230026, China
}%
\affiliation{
Synergetic Innovation Center of Quantum Information \& Quantum Physics, University of Science and Technology of China, Hefei, Anhui 230026, China
}%
\author{Yingwen Zhang}
\affiliation{National Research Council of Canada, 100 Sussex Drive, Ottawa, K1A0R6, Canada
}%
\author{Chen Yang}
\affiliation{Key Laboratory of Quantum Information, University of Science and Technology of China, Hefei, Anhui 230026, China
}%
\affiliation{
Synergetic Innovation Center of Quantum Information \& Quantum Physics, University of Science and Technology of China, Hefei, Anhui 230026, China
}%
\author{Shikai Liu}
\affiliation{Key Laboratory of Quantum Information, University of Science and Technology of China, Hefei, Anhui 230026, China
}%
\affiliation{
Synergetic Innovation Center of Quantum Information \& Quantum Physics, University of Science and Technology of China, Hefei, Anhui 230026, China
}%
\author{Zheng Ge}
\affiliation{Key Laboratory of Quantum Information, University of Science and Technology of China, Hefei, Anhui 230026, China
}%
\affiliation{
Synergetic Innovation Center of Quantum Information \& Quantum Physics, University of Science and Technology of China, Hefei, Anhui 230026, China
}%
\author{Yinhai Li}
\affiliation{Key Laboratory of Quantum Information, University of Science and Technology of China, Hefei, Anhui 230026, China
}%
\affiliation{
Synergetic Innovation Center of Quantum Information \& Quantum Physics, University of Science and Technology of China, Hefei, Anhui 230026, China
}%
\author{Yan Li}
\affiliation{Key Laboratory of Quantum Information, University of Science and Technology of China, Hefei, Anhui 230026, China
}%
\affiliation{
Synergetic Innovation Center of Quantum Information \& Quantum Physics, University of Science and Technology of China, Hefei, Anhui 230026, China
}%
\author{Zhiyuan Zhou}%
\email{zyzhouphy@ustc.edu.cn}
\affiliation{Key Laboratory of Quantum Information, University of Science and Technology of China, Hefei, Anhui 230026, China
}%
\author{Guangcan Guo}
\affiliation{Key Laboratory of Quantum Information, University of Science and Technology of China, Hefei, Anhui 230026, China
}%
\affiliation{
Synergetic Innovation Center of Quantum Information \& Quantum Physics, University of Science and Technology of China, Hefei, Anhui 230026, China
}%
\author{Baosen Shi}
\email{drshi@ustc.edu.cn}
\affiliation{Key Laboratory of Quantum Information, University of Science and Technology of China, Hefei, Anhui 230026, China
}%
\affiliation{
Synergetic Innovation Center of Quantum Information \& Quantum Physics, University of Science and Technology of China, Hefei, Anhui 230026, China
}%

\date{\today}% It is always \today, today,
\begin{abstract}
Photon pair entangled in high dimensional orbital angular momentum (OAM) degree of freedom (DOF) has been widely regarded as a possible source in improving the capacity of quantum information processing. The need for the generation of a high dimensional maximally entangled state in the OAM DOF is therefore much desired. In this work, we demonstrate a simple method to generate a broader and flatter OAM spectrum, i.e. a larger spiral bandwidth (SB), of entangled photon pairs generated through spontaneous parametric down-conversion by modifying the pump beam profile. By investigating both experimentally and theoretically, we have found that an exponential pump profile that is roughly the inverse of the mode profiles of the single-mode fibers used for OAM detection will provide a much larger SB when compared to a Gaussian shaped pump.

\end{abstract}

\pacs{Valid PACS appear here}% PACS, the Physics and Astronomy
\maketitle
\section{Introduction}
Two-photon high dimensional entangled state (HD-ES), $\sum_{j=0}^{d-1}c_j\ket{j}_A \ket{j}_B$, has been widely regarded as useful in increasing capacity for quantum information processing. From the fundamental physic standpoint, such states imply a larger violation of local-realism theories and a lower fidelity bound in quantum state cloning \cite{Dada2011, Collins2002,erhard2018twisted}. Also great attention is given to their practical applications \cite{Yao2011a,erhard2018twisted}. For example,  enhancing security robustness against eavesdrop in quantum cryptography \cite{mirhosseini2015high,cerf2002security}, increasing dimensions of the Bell state in dense coding, entanglement swapping or teleportation \cite{Wang2017,hu2019experimental, LuoYi}, multiplexing heralded single photon source \cite{puigibert2017heralded,liu2019multiplexing},and improving the quality of imaging or quantum sensors \cite{chen2014quantum,zhang2019multidimensional,asban2019quantum}.

In photonic systems, one could construct an HD-ES in many of the photon's degrees of freedom (DOF) \cite{erhard2019advances,forbes2019quantum}. For example, in orbital angular momentum (OAM) \cite{Mair2001,Neves2005,Vaziri2002,Molina-Terriza2007,Zhang2016,walborn2004entanglement,Kovlakov2017}, in paths \cite{Hu2016,Krenn2017,wang2018multidimensional}, frequency \cite{Kues2017}, photon number \cite{Bimbard2010}, or temporal modes \cite{Grassani2015,Li2017}. HD-ES in OAM has been gaining more attention due to their easy scalability in dimension. One typical progress is to create $100 \times 100$ dimensional entanglement via employing both the OAM and radial DOFs of entangled photon pairs \cite{krenn2014generation}.

The most common method in generating OAM entangled photon pairs is via the process of spontaneous parametric down-conversion (SPDC) \cite{Mair2001}. According to OAM conservation, the sum of OAM from the signal and idler photons must equal to that of the pump photon i.e., $\ell_p=\ell_s+\ell_i$. When $\ell_p=0$, the output two-photon state of SPDC can be Schmidt decomposed into $\sum_{-\infty}^{\infty}C_{\ell_s,\ell_i}\ket{-\ell}_s\ket{\ell}_i$. Here, $C_{\ell_s,\ell_i}$ is the probability amplitude($\sum|C_{\ell_s,\ell_i}|^2=1$) of finding the signal photon with OAM $-\ell$ and the idler photon with OAM $\ell$ in coincidence. The width of the OAM spectrum is often known as the spiral bandwidth (SB) \cite{Torres2003}. The Schmidt number $K=1/\sum C_{\ell_s,\ell_i}^4$ is also often defined to evaluate the entanglement dimensions \cite{law2004analysis,miatto2011full,zhang2014modal}; a larger value of $K$ depicts a larger dimensions of entanglement. For a maximally entangled state (MES) of $|C_{\ell_s,\ell_i}|=1\sqrt{d}$, the Schmidt number is $d$. Entangled photons with a larger Schmidt number could be beneficial to implementing higher dimensional quantum protocols like cryptography, computation, imaging and metrology.

For SPDC, the coincidence amplitudes $C_{\ell_s,\ell_i}$  can be calculated via the overlap integral between the input pump mode and both of the signal and idler modes in the Laguerre-Gaussian(LG) basis \cite{miatto2011full, Yao2011}. Previous works have shown several ways to increase the SB of OAM entanglement. Firstly, one can adjust the beam waist ratio between the pump and the measured LG modes $\gamma= w_p/w_{s(i)}$ \cite{miatto2011full,Torres2003,law2004analysis}; the SB increases with increasing $\gamma$. Second is by changing the down-conversion angle through adjusting the SPDC phase matching \cite{romero2012increasing,zhang2014modal,pires2010measurement}; the SB changes based on the down-conversion angle between the SPDC photons. Lastly, is to engineer a crystal with spatially varying phase matching \cite{torres2004quasi,lu2015orbital,hua2018annual}; there has been little experimental progress in this avenue due to the complex fabrication technology. Also, some works attempt to prepare HD-MES  through some complex engineering of the pump beam profile \cite{Torres2003a,Kovlakov2017,machado2019engineering,kovlakov2018quantum,liu2018coherent,shi2020entangled}. In recent work \cite{kovlakov2018quantum,liu2018coherent}, a three dimensional maximally entangled state $1/\sqrt{3}(\ket{-1}\ket{-1}+\ket{0}\ket{0}+\ket{1}\ket{1})$ has been generated by shaping the pump into a superposition of several LG modes. However, the generation of higher dimensional MES remains difficult due to the  crosstalk between OAM modes.

Here, we proposed an ingenious method to increase the SB via some simple shaping of the pump beam profile. We show both theoretically and experimentally that an exponential pump will significantly flatten the OAM spectrum and extend the SB. The optimal exponential pump  is when roughly equal the inverse of the combined profile of the single-mode optical fibers (SMF) for detection. We then performed high dimensional quantum state tomography in a three- and five-dimensional subspace using the optimized exponential pump; the corresponding fidelities are 90.74\% and 81.46\% for three- and five- dimensional MES, respectively. Traditionally, when the input pump beam profile is a Gaussian function, the distribution of coincidence amplitude with OAM is strongly mode-dependent \cite{miatto2011full, Yao2011}. Therefore, mode post-selection has to be performed in order to generate a HD-MES. Our method demonstrates a simple way to broaden and flatten the SB, which could allow future quantum protocols using the OAM DOF to access higher dimensional MES without requiring mode filtering.

 \section{Results}
 \subsection{Optimizing  input beam profile to increase entangled dimension.}
SPDC photons entangled in arbitrary superpositions of OAM are often described in terms of the LG modes. In our analysis, we are only interested in the OAM DOF and set the radial momenta to be zero. In the thin crystal limit, the phase matching function of the SPDC process can be approximated to unity and the coincidence amplitudes $C_{\ell_s,\ell_i}$ can be calculated from the overlap integral \cite{miatto2011full,zhang2014radial,zhang2014simulating}
\begin{equation}
  C_{\ell_s,\ell_i}=\int\Phi(\bm{x})[\text{LG}_{\ell_s}(\bm{x})]^*[\text{LG}_{\ell_i}(\bm{x})]^*\text{G}^2(\bm{x})d^2x,
\label{E1}
\end{equation}
where $\Phi(\bm{x})$ is the mode function of the pump and $\text{G}(\bm{x})$ is the Gaussian mode of the SMF used for detection. $\text{LG}_{\ell}(\bm{x})$ is the LG mode \cite{Allen1992}.
When the pump profile is a Gaussian and by choosing the LG mode size of the signal and idler beams to be equal $w_s=w_i\equiv w_{si}$, the coincidence probability can be evaluated as \cite{miatto2011full,zhang2014radial,zhang2014simulating}
\begin{equation}
	C_{-\ell,\ell}\propto\left(\frac{2\gamma^2}{2\gamma^2+2\eta^2+1}\right)^{|\ell|}.
\label{E2}
\end{equation}
Here $\gamma=w_p/w_{si}$ is the beam waist ratio between that of the pump and the LG modes of the signal and idler photons measured at the nonlinear crystal plane. $\eta=w_p/w_f$ is the beam waist ratio between the pump beam and the mode size of the SMFs. Based on the Eq.(\ref{E2}), the OAM spectrum always peaks at $\ell=0$ and rapidly decreases with increasing $\ell$.
\begin{figure}[htbp]
	\centering
	\includegraphics[width=9cm]{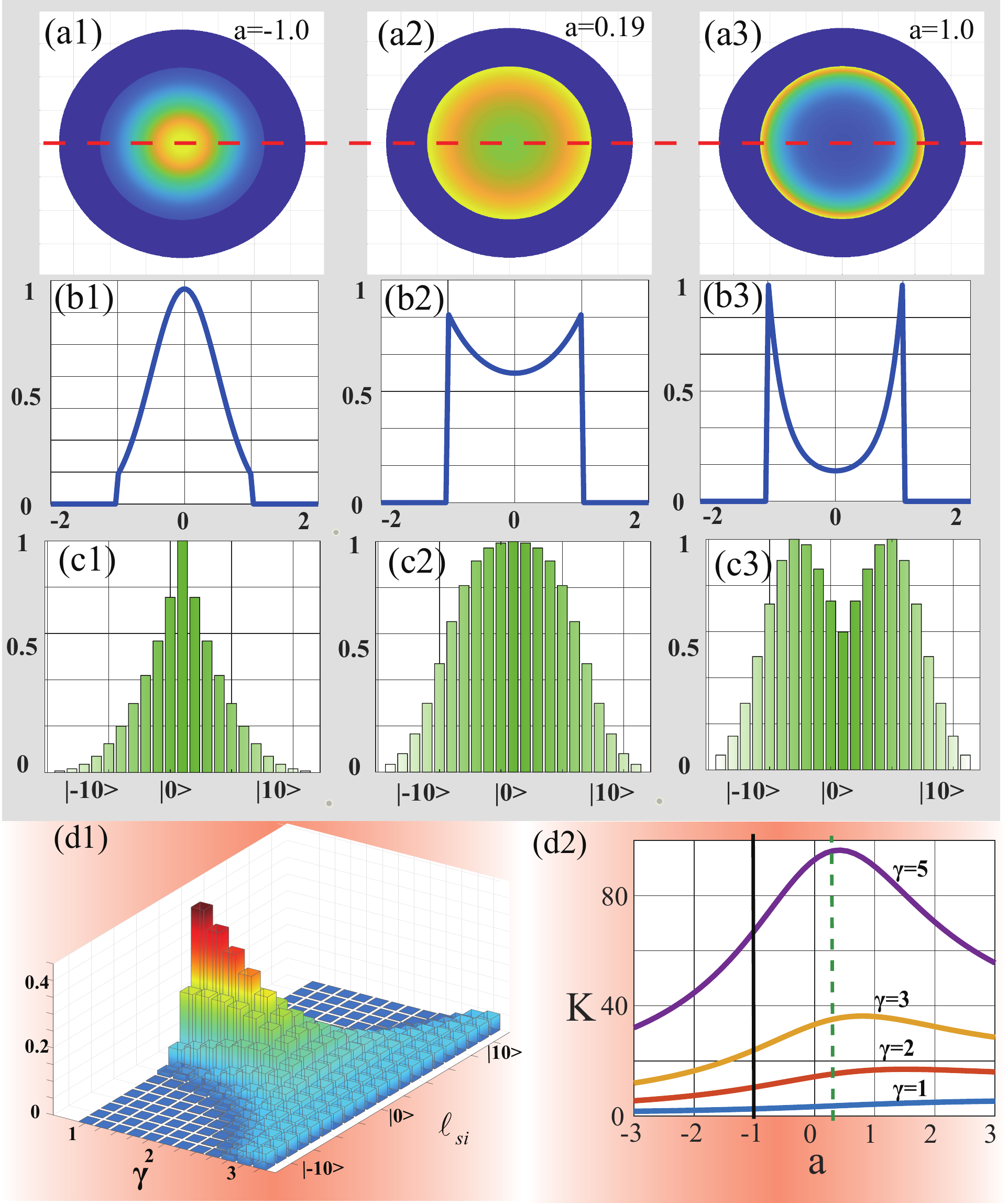}%{Fig_1_V1.eps}
	\caption{Theoretical results in OAM spectrum under  various pump profiles. (a1)-(a3): The intensity pump beam profiles with $a$=-1, $a$=0.19 and $a$=1.0. (b1)-(b3): The corresponding cross-section profile of the pump (along red line in (a1)-(a3)). (c1)-(c3): The theoretically predicted OAM spectrum based on Eq.(\ref{E3}), where $\gamma =w_p/w_{si}$ is equal to 2.0 and $\eta=w_p/w_f$ is equal to 0.31. (d1): The OAM spectrum along different $\gamma$, where the pump beam profile is set to $a=0.19=2\eta^2$. (d2): The Schmidt number versus beam profile parameter of $a$ running from -3 to 3 for various $\gamma$, with OAM ranging from -50 to 50 in the calculations. The left solid  and right dashed vertical  lines in (d2) represent the pump beam profile of a Gaussian $a=-1$ and an exponential with $a=2\eta^2=0.19$, respectively.
}\label{F1}
\end{figure}

Looking at Eq.~(\ref{E1}), we see that if the combined profile (CP) of $\Phi(\bm{x})$ and $\text{G}^2(\bm{x})$ is a constant, the overlap integral should be a constant with respect to $\ell$ resulting in a HD-MES. \cite{zhang2014simulating} classically simulated this using the Klyshko's advanced-wave representation \cite{Klyshko1988}. It was found that the SB can be expanded significantly when this CP is flat. We shall look at a more general situation that uses an adjustable exponentially shaped pump beam whose beam profile is given by
\begin{equation}
	\Phi(r) = \exp\left(\frac{ar^2}{w_p^2}\right)*H(-r + w_p).
\end{equation}

Here $a$ is a parameter that determines the width and curvature of the exponential function and $H(x)$ is the heaviside step function which limits the width of the beam to same. After evaluating the overlap integral in Eq.~(\ref{E1}), $C_{-\ell,\ell}$ is determined to be
\begin{align}
C_{-\ell,\ell} \propto& \left(\frac{2\gamma^2}{2\gamma^2+2\eta^2-a}\right)^{|\ell|}\nonumber\\
&\times \left[1-\frac{1}{|\ell|!}\Gamma\left(1+|\ell|,2\gamma^2+2\eta^2-a\right)\right],
\label{E3}
\end{align}
with $\Gamma(n,z)(=\int_{z}^{\infty}t^{n-1}e^{-t}dt)$ being the incomplete gamma function.

 Eq.~(\ref{E3}) is similar to Eq.~(\ref{E2}) albeit the extra gamma function. Some interesting behaviour can be observed when $a$ is varied:

i) When $a<2\eta^2$, the CP of the pump and the SMFs is still a Gaussian. The OAM spectrum is essentially the same as that in Eq.~(\ref{E2}) (with some small deviations coming from the incomplete gamma function at larger $\ell$) where it peaks at $|\ell|=0$ and decreases rapidly with larger $|\ell|$ values (Fig.~\ref{F1}(a1)). The SB broadens as $a$ approaches $2\eta^2$.

ii) When $a=2\eta^2$, the CP will be a flat-top. The first term in Eq.~(\ref{E3}) becomes a constant resulting in a flat OAM spectrum, however, the incomplete gamma function will suppress $|C_{-\ell,\ell}|^2$ for larger $|\ell|$ values (Fig.~\ref{F1}(a2)). To further broaden the OAM spectrum, one can increase $\gamma$ as seen in Fig.~\ref{F1}(d1).

iii) When $a>2\eta^2$, the CP is an exponential. The denominator in the first term of Eq.~(\ref{E3}) is now smaller than it's numerator, so the term will grow with increasing $|\ell|$. $|C_{-\ell,\ell}|^2$ is still suppressed by the incomplete gamma function at larger $|\ell|$ values. This results in an OAM spectrum that peaks at some non-zero $|\ell|$ value (Fig.~\ref{F1}(a3)).

In Fig.~\ref{F1}(d2), it can be seen that for larger $\gamma$ values, the Schmidt number is a maximum when $a\approx 2\eta^2$. However, when $\gamma$ is small, the maximum Schmidt number occurs at $a>2\eta^2$. This is a result of a larger contribution from the incomplete gamma  function when $\gamma$ is small therefore suppressing $C_{-\ell,\ell}$  at smaller $\ell$ values.

\subsection{Beam shaping technology for two-photon high dimensional entanglements}
\begin{figure}[ht]
	\centering
	\includegraphics[width=9cm]{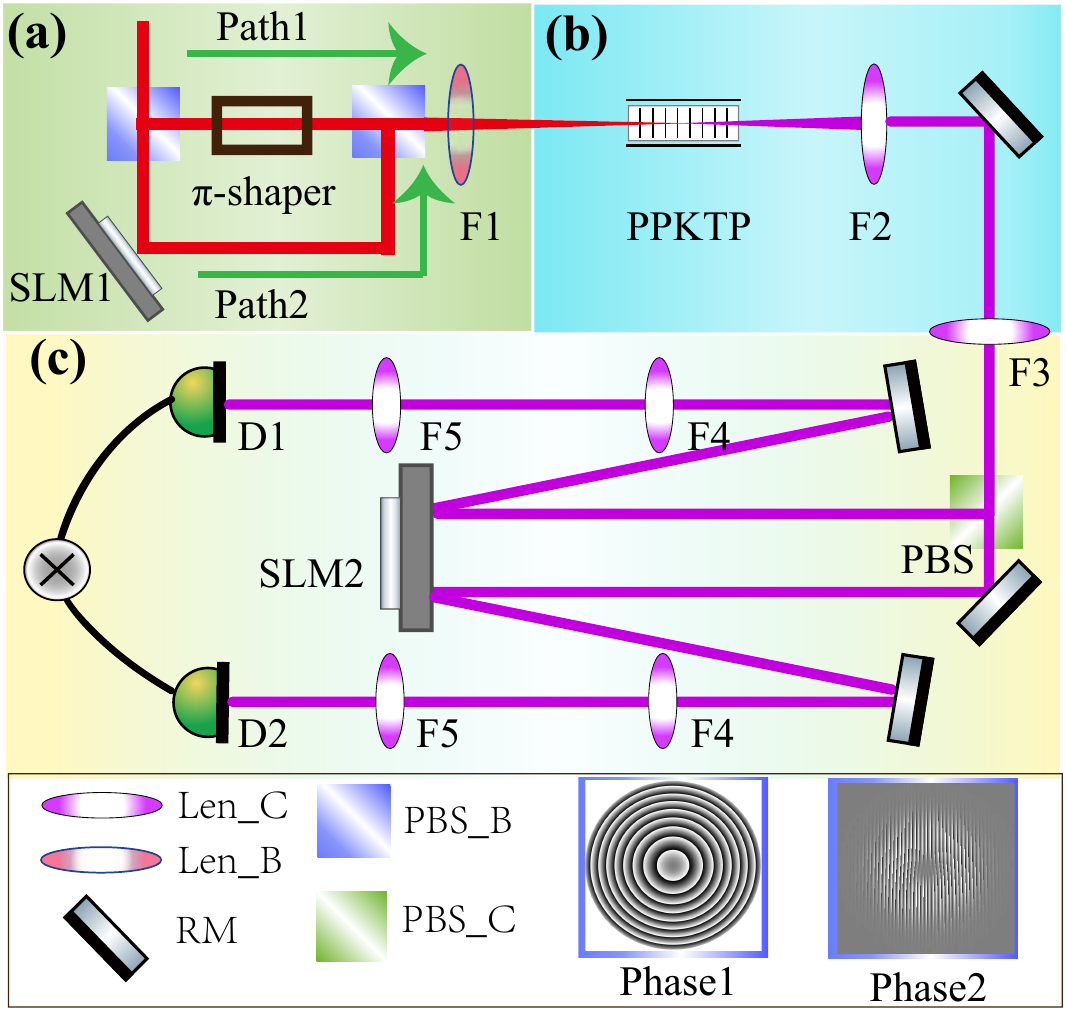}%{Fig_2_V1.eps}%{Fig_2_V1.eps}
	\caption{Experimental setup for pump shaping, generation and detection of OAM entangled photons. The pump beam is first shaped either by SLM1(i.e.,Phase1 for $a=0$) or the $\pi$-shaper into the desired beam shape. The beam then pumps, a 10mm long PPKTP crystal (type-II, 775um$->$1550um) to generate OAM entangled photon pairs via SPDC. Through the use of SLM2 (i.e., Phase2 for $\ell=1$) and single-mode fibers, projective measurements of the photons' OAM can be performed.}
\label{F2}
\end{figure}

To verify the theoretical prediction in Eq.~(\ref{E3}), we measured two-photon OAM correlations using different input beam profiles (parameter $a$). The corresponding experimental setup is shown in Fig.~\ref{F2}, which includes three parts: pump beam shaping - Fig.~\ref{F2}(a), state generations - Fig.~\ref{F2}(b), and projection measurements - Fig.~\ref{F2}(c). First, the pump beam is shaped using either a SLM (Path2) or a $\pi$-shaper (Path1) from a Gaussian into the desired beam shape (details in the appendix A). Then, a 10mm long nonlinear crystal (PPKTP) is placed the beam waist to perform the SPDC process. The spatial photons at the nonlinear crystal plane was then imaged to the surface of another SLM for mode demodulation and then for coincidence measurements via  a superconducting nanowire single-photon detector. In our setup, the beam width ratio $\eta$ is 0.31. It should be noted that though a SLM have more versatility in the beam shaping it can perform, however, it could not support high pump intensities and have lower conversion efficiencies compared to a commercial $\pi$-shaper. The SLM is therefore used in confirming the shape of the SB for various $a$ parameters, in situations where high pump power (350mW) is required to increase SPDC photon production rate and reduce data acquisition time without needing to change the beam shape, the $\pi$-shaper is used.

\subsection{Two-photon OAM spectrum under the different beam profiles}

\begin{figure}[!ht]
  \centering
  \includegraphics[width=9cm]{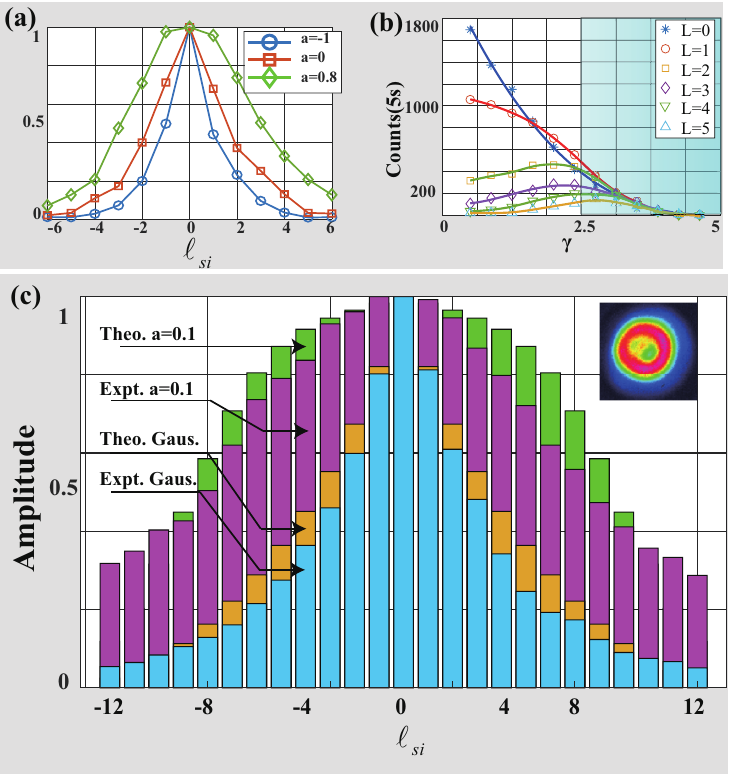}%{Fig_3_V2.eps}
 \caption{The normalized OAM spectrum in SPDC. (a): OAM spectrum for various pump profiles. Here, $\gamma=1.25$ and $\eta=0.31$. (b): The coincidences for various $\ell$ values as a function of $\gamma$ when the pump profile has $a$=0.10. (c): an OAM spectrum measured from $\ell$=-12 to 12 for $\gamma=2.4$, the green(the fourth layer,innermost) and red(the third layer)  OAM spectrum are normalized theoretical and experimental results for $a$=0.10, respectively; the corresponding Gaussian $a$=-1 are shown in orange(the second layer) and blue bars(the first layer, outmost). The inset on the top right of (c) is the optimized pump beam profile, where we can estimate the parameter a=0.10 by fitting it's intensity. The non-normalized versions of (a) and (c) can be found in the appendix A.}\label{F3}
\end{figure}

Fig.~\ref{F3}(a) shows the OAM spectrum generated with different pump beam profiles ($a=-1$, 0, and 0.8)that can be attached in appendix A, where the beam width ratio $\gamma$ is 1.25. It can see that the OAM spectrum broadens as $a$ increases, just as theoretically predicted from Eq.~(\ref{E3}). Also to note since $\gamma$ is small, the optimized beam profile $a$ ( the largest SB (or $K$)) is actually larger than $2\eta^2$ as seen in Fig.~\ref{F1}(d2). The theoretical OAM spectrum for these three cases can be found in the appendix A.

In addition to the beam profile parameter $a$, the beam width ratio $\gamma$ is another important parameter that affects the entangled dimension.  In Fig.~\ref{F3}(b), we show the measured coincidence rate for various $\ell$ values as a function of $\gamma$, where $a$=0.10. One can see that for larger value of $\gamma$ (blue area in Fig.~\ref{F3}(b) $\gamma>=2.5$), the difference in coincidence rate for the various $\ell$  are relatively small which indicates a broader SB. This is in agreement with our theoretical results shown in Fig.~\ref{F1}(d1).

% 我们也研究了更多的模式情况。、、、
In Fig.~\ref{F3}(c) we show the theoretical and experimental OAM spectrum from $\ell=-12$ to 12 for $\gamma=2.4$. A significantly broader OAM spectrum can be observed when compared to a Gaussian pump ($a=-1$) with the same $\gamma$. The experimental azimuthal Schmidt number $K$ is determined to be 21.9, which is in good agreement with the theoretical prediction of 20.7, for a Gaussian pump $K$ is 15. For quantifying the crosstalk between two neighboring OAM values, we measured the crosstalk-visibility $(1-\sum_{i,j=i\pm 1}{C_{i,j}^2}/\sum_{i=-12}^{12}C_{ii}^2)$ and obtained a value of 93.91\%.

\subsection{Quantum state tomography of  high dimensional entanglement}
\begin{figure}[htbp]
 \centering
 \includegraphics[width=9cm]{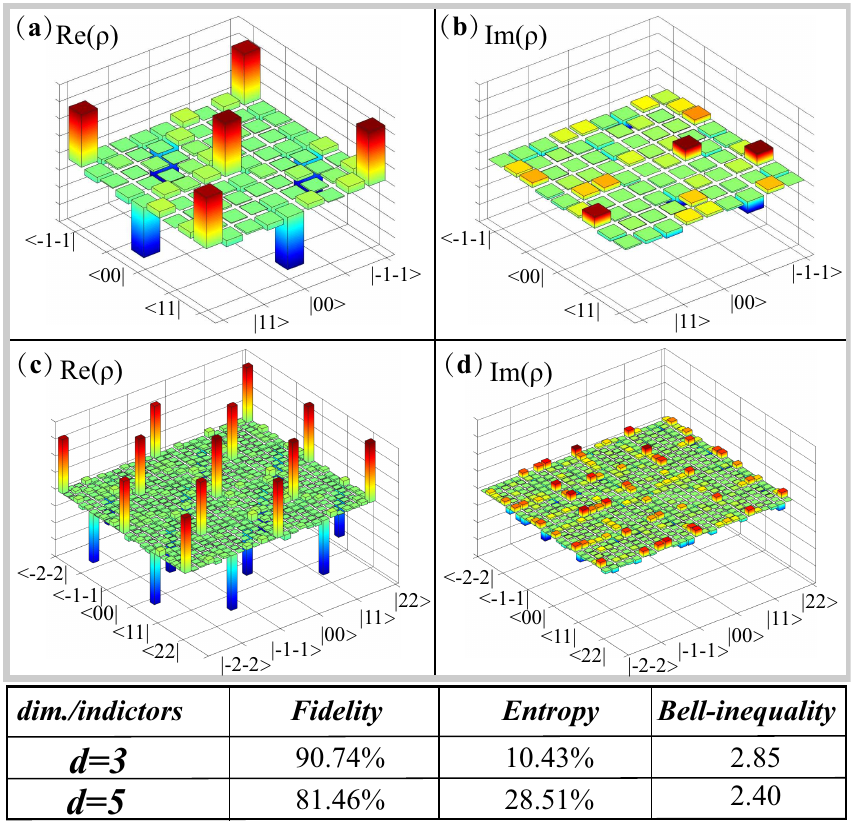}%{Fig_4.eps}
 \caption{The reconstructed  density matrices of three and five dimensional OAM MES from the experimental OAM spectrum in Fig.~\ref{F3}(c). (a) and (b): The real and imaginary parts of the density matrix for $d=3$. (c) and (d): The real and imaginary parts of the density matrix for $d=5$. The table under (c) and (d) displays the corresponding fidelity, entropy, and CGLMP-Bell inequality for the two density matrices. The data acquisition times for each reading is 50s and 300s for the three and five dimensional cases respectively.}
\label{F4}
\end{figure}
From Fig.~\ref{F3}(c), one can see that the HD-MES is prepared at least in a five-dimensional subspace. We reconstructed the density matrices for the cases of dimension $d=3$ (Fig.~\ref{F4}(a), (b)) and $d=5$ (Fig.~\ref{F4}(c), (d)) through high dimensional quantum state tomography \cite{Thew2002_QST,Giovannini2013,liu2018coherent}(also see appendix B). The measured fidelity, $F=[Tr\sqrt{\sqrt{\rho}\rho_{\text{exp}}\sqrt{\rho}}]^2$, was $0.9071\pm0.005$ for $d=3$ and $0.8146\pm0.0014$ for $d=5$ with the uncertainty obtained through statistical simulations that assumed the coincidence events follow a Poissonian distribution. The fidelity for both $d=3$ and $d=5$ entangled states exceeded the dimensional threshold of $(d-1)/d$, signifying that the density matrix cannot be decomposed into an ensemble of pure states with low Schmidt number \cite{sanpera2001schmidt,bavaresco2018measurements,friis2019entanglement}. The fact that the fidelity of the $d=5$ MES is less than that of $d=3$ is not surprising as when we look at Fig.~\ref{F3}(c), the OAM spectrum is fairly flat at $d=3$ ($\ell =$ -1, 0, 1), but less so at $d=5$ ($\ell =$ -2, -1, 0, 1 2). One can improve the fidelity for higher dimensions by using a slightly larger beam waist ratio $\gamma$.  From the density matrix we can also calculate the linear entropy, $S_{ent}=1-Tr(\rho_{exp}^2)$, giving $S_{ent}=0.1043\pm0.009$ and  $0.2851\pm 0.0093$ for the $d=3$ and $d=5$ cases respectively, the linear entropy determined using the theoretical OAM spectrum is $4*10^{-4}$ for $d=3$ and $12*10^{-4}$ for $d=5$ (for a pure state the linear entropy is zero). Furthermore, the CGLMP Bell inequality \cite{Collins2002} was determined to be $2.85\pm 0.03$ and $2.40\pm 0.01$ for the $d=3$ and $d=5$ entangled states respectively. As a comparison, the theoretical upper bound for the CGLMP Bell inequality is 2.87 and 2.91 for the $d=3$ and $d=5$ entangled states respectively\cite{Collins2002}. The lower values for the $d=5$ case is mainly attributed to imperfect mode overlap between the SPDC photons and the measurement SLM. These values are listed in the table below Fig.~\ref{F4} for clarify.

\section{Discussion}
In this work, a simple technique of shaping the pump beam profiles to increase the two-photon OAM entanglement dimensions in a SPDC processes is demonstrated. Theoretically and experimentally, we found that the coincidence amplitude will become mostly mode independent when the pump profile is an exponential that roughly cancels the Gaussian profile of the SMFs used for photon detection. When compared to the more commonly used method of increasing $\gamma$ to expand the OAM entanglement dimensions, optimizing the beam profile in SPDC offers two advantages. Firstly, by optimizing the beam profile one would not suffer much losses to the coincidence count rate as compared to increasing $\gamma$, which could lead to significant reduction in the coincidence count rate due to a decrease in the coupling efficiency into the SMFs. Secondly, by shaping the pump into an inverse Gaussian, one could always achieve an OAM spectrum that is flat for at least several OAM modes even in cases when $\gamma$ is small ($\approx 2$ or less), this allows one to generate a HD-MES in situations not possible previously. This adds a new way of expanding the SPDC OAM spectrum and can be used concurrently with previously suggested techniques such as increasing $\gamma$ and adjusting the down conversion angle between the signal and idler photons. The ability to generate such HD-MES without mode post-selection will be of great importance in quantum communication, quantum sensing and also in fundamental physics research.

\section{Acknowledgments}
 This work is supported by The Anhui Initiative in Quantum Information Technologies (AHY020200); National Natural Science Foundation of China (61435011, 61525504, 61605194,11934013); China Postdoctoral Science Foundation (2016M590570, 2017M622003); Fundamental Research Funds for the Central Universities.
%\nocite{bibitem1}

\appendix

\section{Shaping the beam profile via diffractive optics}
 When a monochrome electronic field $U(x_1,y_1)$ passes a lossless phase elements, the mapping of the output field can be expressed in terms of Fresnel integral. Usually, a lens ($f$) is used to focus the beam waist located after the phase element. The beam shaping problem can be seen as a Fourier transform \cite{dickey2018laser,rosales2017shape}:
\begin{equation}\label{S5}
\begin{array}{l}
U(x,y) = \frac{1}{{i\lambda f}}\exp (ikf + {x^2} + {y^2}) \times \int {\int U } ({x_1},{y_1})\\
 \times \exp (i\beta \phi ({x_1},{y_1}))exp( - i\frac{{2\pi }}{{\lambda f}}(x{x_1} + y{y_1}))d{x_1}d{y_1}
\end{array}
\end{equation}
where $k=2\pi/\lambda$, $U(x_1,y_1)$ and $U(x,y)$ are magnitudes of the input and output fields, respectively. The beam shaping problem is to determine what the phase function $\beta \phi(x_1,y_1)$ is. Here, $\beta(=2\pi w_0w_1/\lambda f)$ is a system parameter connecting the input beam width $w_0$ with output the beam width $w_1$. In principle, any arbitrary field can be shaped approximately through a suitable phase function. A suitably large $\beta$ can generate a good approximate output field\cite{dickey2018laser}.

Based on the diffraction theory of lossless beam shaping \cite{dickey2018laser}, three steps are needed to determine the phase factor $\phi$. For a radially symmetric problem, first, we need to evaluate the constant $A$ given by
\begin{equation}\label{S6}
A = \frac{{\int\limits_{ - \infty }^{ + \infty } {I(s)ds} }}{{\int\limits_{ - \infty }^{ + \infty } {Q(s)ds} }}
\end{equation}
where the $I(s)$ and $Q(s)$ are the intensities of the input and output beams. Then, the phase factor $\phi$ can be determined by solving two ordinary differential equations (ODE) \cite{dickey2018laser}:
\begin{equation}\label{S7}
  \left\{ \begin{array}{l}
AQ(\alpha ) \cdot \frac{{d\alpha }}{{d\xi }} = I(\xi )\\
\frac{{d\phi }}{{d\xi }} = \alpha (\xi ),
\end{array} \right.
\end{equation}
where $\alpha(\xi)$ is a medium function. For some certain special patterns, the ODE can be solved analytically, for example, for the  output as a flat-top beam $(Q(s)=1*H(-s+1))$, the phase could be given:
\begin{equation}\label{4}
\phi (\xi ) =  - \frac{2}{\pi }\left( {\xi \frac{{\sqrt \pi  }}{2}\exp (\xi ) + \frac{1}{2}\exp ( - {\xi ^2}) - \frac{1}{2}} \right)
\end{equation}
In most situations, one needs to solve for $\phi$ numerically, which is the case when the output beam profile is an exponential function $Q(s)=\exp(as^2)$ as in the paper.

\begin{figure}[htb]
  \centering
  \includegraphics[width=9cm]{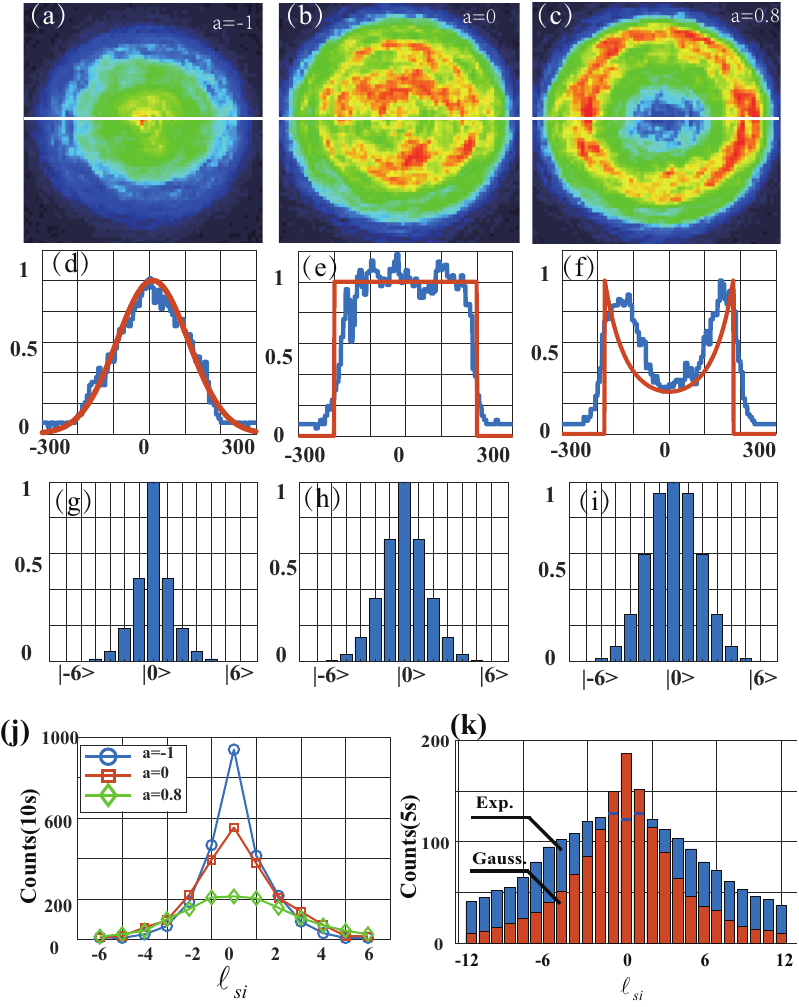}%{Sup_Fig_1.eps}
  \caption{The intensity patterns of pump beam profiles and the corresponding OAM spectrum. (a)-(c): The pump beams generated in our system with $a=-1$, 0, and 0.8, respectively. (d)-(f): The corresponding cross-section intensity profile (along the white line through (a)-(c)). The blue winding lines are the measured distributions, and the red smooth lines are the theoretical distributions. (g)-(i): The corresponding theoretical OAM spectrum  with $\gamma=1.25$. (j)-(k): The experimental OAM spectrum corresponding to Fig. ~\ref{F3}(a) and (c), respectively. In ~\ref{F3}(c), the outmost red distribution is the situation of Gaussian pump, and the innermost blue bars is the situation of optimized  exponential pump}\label{F5}
\end{figure}
Experimentally, we can realize the beam shaping by loading the phase $\beta \phi$ onto a SLM. In our setup (Fig. 2 in the main text), the Fourier lens (F1) has a focal length of 75 mm; the input beam waist is 3000 um; the beam width of the output beam is set to 200um in the Fourier plane. Therefore, one can determine $\beta$=64.4. Fig. ~\ref{F5}(a)-(c) shows some intensity patterns of the input beams with the exponential function. The corresponding cross-section distributions are shown in Fig. ~\ref{F5}(d)-(f). Under these beam as a pump, we get the theoretical OAM spectrum, which is shown in Fig. ~\ref{F5}(g)-(f); Fig. ~\ref{F5}(j) is the corresponding experimental results. Fig. ~\ref{F5}(k) is the measured OAM spectrum when the situation of the pumps are the optimized and Gaussian beam, respectively, where we use the $\pi-$shaper to generate the optimized pump beam.

The used SLM could not support very high input power and has a low conversion efficiency (about 30\% in first-order). Therefore, we employed a commercial $\pi$-shaper to perform beam shaping when the high pump power is required to increase SPDC photon production and reduce data acquisition time. In this regime, a $\pi-$shaper and a Fourier lens($f=1000mm$) are used to shape the beam profile. A $\pi-$shaper is used to transform a Gaussian into an Airy disk via the Fourier-Bessel transformation \cite{laskin2013beam,liu2020highQFC}:
\begin{equation}\label{App-A1}
  {I_f}(\rho ) = {I_{f0}}{\left[ {{J_0}(2\pi \rho )/2\pi \rho } \right]^2},
\end{equation}
where ${J_0}(2\pi \rho )$ is the zeroth-order Bessel function of the first kind and $I_{f0}$ is the normalization factor. Such a beam can be transformed into a flat-top beam in the Fourier plane by using a Fourier lens. The change in the shape of $I(\rho)$ as it propagates can be determined by the Rayleigh-Sommerfeld diffraction integral \cite{brosseau1998fundamentals}. An exponential-like beam shape would be created at a location slightly away from the Fourier plane. Experimentally, we could move the nonlinear crystal along the beam axis slightly to the location with the desired beam shape.

   \section{The details of high dimensional quantum entangled state tomography}
   For a high dimensional entangled state (MES), the theoretical density matrices can be given as:
   \begin{equation}
     \rho=\ket{\psi}_{MES}\otimes\bra{\psi}_{MES}
   \end{equation}
   Using this definition, we can calculate the theoretical density matrix $\rho$ of MES \cite{Giovannini2013}.

Experimentally, with the help of projection- measurement, one  could reconstruct the density matrix of high dimensional MES. For an HD-MES defined in $d$ dimensional space, the corresponding reconstructed density matrices can be written as\cite{Thew2002_QST}:
\begin{equation}\label{App-B0}
  \rho_{exp}=N\sum_{u,v,j,k=1}^{d^2}(A_{uv}^{jk})^{-1}n_{uv}\lambda_j
  \otimes\lambda_k.
\end{equation}
where $N$ is the normalized coefficient; $A_{uv}^{jk}(=\bra{\Psi_{uv}}\lambda_j\otimes\lambda_k\ket{\Psi_{uv}})$ is the constant matrix associating with the fundamental matrix $\lambda_{j,k}$ and measurement basis  $\ket{\Psi_{uv}}$, in which, $\lambda_{j,k}$ can be generated by SU(d) algebra; $\ket{\Psi_{u,v}}=\ket{\Psi_u}_A\ket{\Psi_v}_B^{\dag}$ represents the measurement basis in signal (A) and idler photons(B); $(A_{uv}^{jk})^{-1}$ are the corresponding inverted matrices; $n_{uv}=Ntr(\Pi_{A,B}\rho_{exp})$ represents the coincidence counts measured by electronic systems \cite{Giovannini2013}. In order to experimentally reconstruct density matrix  of HD-MES, three steps should be performed in following.

1.  Ensure the details of the projection- measurement basis.

The first step is to ensure the details of the projection- measurement basis.
 The constant matrix $A_{uv}^{jk}$ is associating with the measurement basis  $\ket{\Psi_{uv}}$. One could set the measurement basis to a complete group of mutually unbiased bases (MUBs) $\left\{ {\left| {\Psi_m^j} \right\rangle } \right\}$, which can be generated using the Weyl group, Hadamard matrix, or Fourier-Gauss transform methods\cite{durt2010mutually}. Here, we used the discrete Fourier-Gauss transform to product MUBs in prime dimensional space \cite{wiesniak2011},
\begin{equation}\label{App-B1}
  \left\{ {{\rm{|\Psi}}_m^j{\rm{ > }}} \right\} = \left\{ {\frac{1}{{\sqrt d }}\sum\limits_{n = 0}^{d - 1} {\omega _d^{\left( {j{n^2} + nm} \right)}} \left| n \right\rangle } \right\}
\end{equation}
Where $j(j=0...d-1)$ indexes the group of the MUBs; $m(m=0...d-1)$ indexes the superposed OAM states for each set in MUBs, and ${\left| {\left\langle {\Psi_m^j} \right|\left. {\Psi_{m'}^{j'}} \right\rangle } \right|^2} = 1/d\left( {1 - {\delta _{jj'}}} \right)$ for the MUBs. In actuality, $j$ runs from 0 to $d$, with the last set of MUBs being the OAM eigenstates.

2. Obtain a series of coincidence photon counts.

When we set the details of MUB, one next perform projection-measurement to get coincidence photon counts $n_{uv}$ under these MUBs, or named OAM superposition states. Experimentally, we employ the amplitude- encoding technology to generate high fidelity OAM-MUBs\cite{bolduc2013exact,liu2019classical_cat}.

3. Calculate experimental density matrix.

 Finally, based on the Eq. ~\ref{App-B0}, we calculate the density matrix $\rho_{exp}$, and thus get the fidelity and entropy. It should be noted that the reconstructed density matrix maybe not a 'physical' density matrix, i.e., it has the property of positive semi-definiteness \cite{James2001}. For overcoming the disadvantage, the maximum likelihood estimation method is used during the process of reconstructions. We build the likelihood function:
\begin{equation}
  L(t_1,t_2...,t_{d^4})=\sum_{j=1}^{d^4}\frac{[N(\bra{\Psi_j}\rho_{exp}\ket{\Psi}_j-n_j]^2}{2N(\bra{\Psi_j}\rho_{exp}\ket{\Psi}_j}
\end{equation}
Where the $\rho_{exp}$ should be preliminary defined a  'physical' density matrix \cite{James2001}.

%\nocite{*}
%\bibliography{5D_MES}% Produces the bibliography via BibTeX.

\end{document}